\documentclass[aps,pra,notitlepage,twocolumn]{revtex4-1}
\usepackage{amsmath,amssymb,graphicx,color}
\usepackage[english]{babel}
\usepackage{yfonts}

\begin{document}

\newcommand{\bra}[1]    {\langle #1|}
\newcommand{\ket}[1]    {\left| #1 \right\rangle}
\newcommand{\braket}[2]    {\left\langle #1 | #2 \right\rangle}
\newcommand{\ketbra}[2]{\left|#1\right\rangle\!\left\langle#2\right|}
\newcommand{\braketbig}[2]    {\big\langle #1 \big| #2 \big\rangle}
\newcommand{\braketbigg}[2]    {\bigg\langle #1 \bigg| #2 \bigg\rangle}
\newcommand{\tr}[1]    {{\rm Tr}\left[ #1 \right]}
\newcommand{\av}[1]    {\langle{#1}\rangle}
\newcommand{\x}{\mathbf{r}}
\newcommand{\bk}{\mathbf{k}}
\newcommand{\bp}{\mathbf{p}}
\newcommand{\re}[1]{\textfrak{Re}\left[#1\right]}
\newcommand{\im}[1]{\textfrak{Im}\left[#1\right]}
\newcommand{\Hz}{\mathrm{Hz}}
\newcommand{\erfi}{\mathrm{erfi}}
\newcommand{\erf}{\mathrm{erf}}
\newcommand{\sinc}{\mathrm{sinc}}
\newcommand{\nn}{\nonumber}

\title{Cooperatively-enhanced precision of hybrid light-matter sensors}
\author{A. Niezgoda$^1$, J. Chwede\'nczuk$^1$, T. Wasak$^2$ and F. Piazza$^2$}
\affiliation{$^1$Faculty of Physics, University of Warsaw, ul. Pasteura 5, PL--02--093 Warszawa, Poland\\
  $^2$Max-Planck-Institut f\"ur Physik komplexer Systeme, 01187 Dresden, Germany}
\begin{abstract}
  We consider a hybrid system of matter and light as a sensing device and quantify the role of cooperative effects. The latter generically enhance the precision with
  which modifications of the effective light-matter coupling constant can be measured. In particular, considering a fundamental model of $N$ qubits coupled to a single
  electromagnetic mode, we show that the ultimate bound for the precision shows double-Heisenberg scaling: $\Delta\theta\propto1/(Nn)$, with $N$ and $n$ being the number
  of qubits and photons, respectively. Moreover, even using classical
  states and measuring only one subsystem, a
  Heisenberg-times-shot-noise scaling, i.e.
  $1/(N\sqrt{n})$ or $1/(n\sqrt{N})$, is reached.  As an application, we show that a Bose-Einstein condensate trapped in a double-well potential within an optical cavity can
  detect the gravitational acceleration $g$ with the relative precision of $\Delta g/g\simeq10^{-9}\text{Hz}^{-1/2}$.  The analytical approach presented in this study
  takes into account the leakage of photons through the cavity mirrors,
  and allows to determine the sensitivity when $g$ is inferred via measurements on atoms
  or photons.
\end{abstract}
\maketitle

\section{Introduction}
The use of hybrid light-matter systems has a large potential for the development of classical and quantum technologies.  The idea of exploiting the best of both worlds
culminates in the concept of a quantum network~\cite{kimble_review,rempe_network,cirac_network}, where photons act as information carriers channeling between nodes, where
the matter is used for information storage and as source of the nonlinearities needed for information processing.  These optical nonlinearities correlate matter with light,
allowing to gain information and even modify the former by measuring the latter.  This permits for instance to control the motion of mechanical objects via light in
optomechanical systems~\cite{marq_rev,meystre_rev}, with important consequences for interferometry of displacement
measurements~\cite{caves_1980,Meystre:85,braginsky_rev,schoelkkopf_rev,kippenberg_displ_exp_2009,schwab_displ_exp_2014}.

For such schemes it is crucial to reach a strong light-matter coupling, which can be achieved by employing optical resonators.  Among the most promising kinds of matter,
neutral atoms stand out due to the high control achievable over internal and external degrees of freedom~\cite{kimble_1998,rbh_qed_rmp,mabuchi_2002}.  For instance,
atom-light coupling can be exploited to efficiently create entanglement in atomic
ensembles~\cite{hald1999spin,kasevich_2008,polzik_rev_2010,vuletic_QND_2010,thompson_QND_2011,meschede_2012,reichel_QND_2014}, which constitutes an alternative route to
the use of intrinsic atom-atom nonlinearities~\cite{esteve2008squeezing, gross2010nonlinear,riedel2010atom,berrada2013integrated,collision_paris,lucke2011twin, twin_beam,
  cauchy_paris,twin_paris,smerzi_ob,shin2019bell}, with applications for quantum metrology beating the shot-noise limit~\cite{giovannetti2004quantum}.  Hybrid devices
exploiting atom-light nonlinearities and cooperative effects for metrology and sensing include white-light interferometers with anomalous
dispersion~\cite{fleischauer_1997,shahriar_whitecav_exp_2007}, superradiance~\cite{scully_SR_metr_2014} and superradiant
lasers~\cite{thompson_SR_magn_2012,thompson_SR_2013}, single-atom cavity-QED platforms for nonclassical light \cite{gietka_cav_noncl_2017}, quantum state-transfer
protocols with information recycling~\cite{haine_rec_2013,haine_hybrid_2014,haine_caves_rec_2015,haine_rec_cavity_2015}, optical
magnetometers~\cite{optical_magn_rev_2007,mitchell_sq_2011} and their nonlinear version~\cite{mitchell_2011}.  In particular, in the field of inertial sensing with
atoms~\cite{berman1997atom,cronin_2009,tinoetal_2010}, the use of optical resonators has been shown to enhance the precision of a Mach-Zehnder
interferometer~\cite{mueller_atomint_cavity_2015} and is for instance expected to improve the sensitivity of Bloch-oscillation-based
metrology~\cite{holland_bloch_cav_2009,odell_bloch_cav_2014}. More
recently, the supersolid phase of ultracold bosons induced by the coupling to an optical resonator
has been predicted to allow for very precise gravimetry \cite{miv_ss_cav_2018,gietka_ss_grav_2019}. Recently, an optical cavity-QED setting with strong cooperative
  atom-light interactions has been used to create nonclassical states
  of light, which allow for electric-field sensing beyond the standard quantum
  limit~\cite{rey2020protocol}. Despite these various applications, a
  systematic study of the performance of hybrid light-matter systems
  is still lacking in the regime where cooperative effects are dominant.

In this work, we characterize the different working regimes of a hybrid light-matter sensor aiming at measuring modifications of the effective light-matter coupling
constant. We consider a minimal model for cooperative effects, consisting of $N$ qubits coupled to a single electromagnetic mode. This model allows for closed analytical
expressions for the measurement error, also called the precision or the sensitivity. We find that the ultimate bound for the error satisfies a double-Heisenberg scaling:
$\Delta\theta\propto 1/(Nn)$, with both the number of qubits $N$ and of photons $n$. We also study the dependence on different initial states (classical and non-classical)
of the system, as well as on different measurements. Even for classical states of qubits and photons, and by simply measuring a qubit or a photon
observable, the error scales partially at the Heisenberg limit, i.e., $\Delta\theta\propto 1/(\sqrt{N}n)$ or $\Delta\theta\propto 1/(N\sqrt{n})$, respectively.

Finally, we consider a specific example where an atomic Bose-Einstein condendsate (BEC) trapped in a double-well potential is dispersively coupled to a single mode of an
optical cavity.  The gravitational acceleration $g$ modifies the
effective atom-photon coupling and this effect is amplified by the
cooperative effects.
We determine the dynamics of the system and analytically calculate the precision assuming that $g$ is deduced either from the homodyne detection of the mean of the
quadrature of light or from the mean imbalance between the atomic occupation of each well.
We show that the
relative error $\Delta g/g$, which scales inversely both with the numbers of atoms and photons, can reach the level of $10^{-9}\text{Hz}^{-1/2}$ 
with realistic parameters and classical states of matter and
light, also including the effect of photon loss.  This precision is
comparable with the one predicted for a supersolid state of atoms in
cavities \cite{gietka_ss_grav_2019}.
Our results can be easily extended to other input states, regimes of
parameters or estimation protocols.

The paper is organized as follows: In Section \ref{sec:bounds} we
introduce the model and derive the ultimate bounds for the
sensitivity, as well as specific bounds for certain types of
measurements and input states.
In Section~\ref{sec:pump-loss} we consider a specific scheme where the
electromagnetic field is coherently driven and lossy and the qubits
are prepared in a Gaussian state.
In Section~\ref{sec.app} we present application of our model in gravity sensing and its possible precision using coherent atomic states. 
We conclude in Section~\ref{sec.conc}. Detailed
analytical calculations are presented in the Appendix.

\section{Model and general precision bounds}
\label{sec:bounds}

In order to demonstrate how cooperative effects can enhance the
sensitivity of a hybrid light-matter sensor we consider a minimal
model describing $N$ qubits all equally coupled to a single mode of
an electromagnetic field, corresponding to the following Hamiltonian {(for the details see Ref.~\cite{szirmai} and Appendix~\ref{app.ham})}
\begin{align}\label{eq.ham}
  \hat H&=(-\Delta_c+a_1N)\hat n+\eta(\hat a+\hat a^\dagger)+a_2\hat n\hat J_x,
\end{align}
where in the rotating frame $\Delta_c$ is the characteristic frequency of the
electromagnetic mode which is coherently driven with a strength
$\eta$, $\hat{n}=\hat{a}^\dag\hat{a}$ is the number of photons in the
mode, and $\hat{J_x}=\frac12\sum_{i=1}^N\hat\sigma_x^{(i)}$ is the $x$-component
of the collective spin operator ($\hat\sigma_x^{(i)}$ is the $x$-axis Pauli matrix for the $i$-th qubit). The Hamiltonian \eqref{eq.ham}
contains two types of light-matter coupling: a static collective shift of the
electromagnetic mode frequency quantified by the coupling constant
$a_1$, and a cavity-induced ``quantized effective magnetic field''  coupled to the collective
spin operator (or, equivalently, a qubit-induced dynamical shift of the mode frequency) with characteristic strength $a_2$.

\subsection{Ultimate bounds on the sensitivity}

We now demonstrate that the system governed by the
Hamiltonian~\eqref{eq.ham} can be employed as a sensor for the
estimation of a parameter $\theta$ entering the light-matter coupling constants
$a_1$ and/or $a_2$, with the best possible precision showing the
double-Heisenberg scaling $\Delta\theta\propto n^{-1}N^{-1}$, where
$n=\langle\hat{n}\rangle$ is the number of photons.

To this end, we recall that 
according to the Cramer-Rao lower bound~\cite{holevo2011probabilistic}, the sensitivity in estimating the value of $\theta$ is bounded from below by
\begin{align}
  \Delta\theta\geqslant\frac1{\sqrt{F_Q}}.
\end{align}
The $F_Q$ is the quantum Fisher information
(QFI)~\cite{braunstein1994statistical} given by
\begin{align}\label{eq.qfi.mix}
  F_Q=\sum_{i,j}\frac{(\lambda_i-\lambda_j)^2}{\lambda_i+\lambda_j}\left|\bra i\hat h\ket j\right|^2,
\end{align}
where $\ket i$'s and $\lambda$'s are the eigenvectors and the corresponding eigenvalues of the density matrix, i.e., $\hat\varrho=\sum_{i}\lambda_i\ketbra{i}{i}$. 
For pure states, when only one $\lambda$ is non-zero, this simplifies to
\begin{align}\label{eq.qfi.pure}
  F_Q=4(\av{\hat h^2}-\av{\hat h}^2)\equiv4\av{(\Delta\hat h)^2}.
\end{align}
The operator $\hat h$ generates the transformation in the parameter space, namely $\hat h=i(\partial_\theta\hat U)\hat U^\dagger$, where $\hat U = e^{-i \hat H t}$ is
  the evolution operator determined by the Hamiltonian from Eq.~\eqref{eq.ham}. It can be rewritten in a more useful form (see Appendix~\ref{app.evolution}), namely
\begin{align}\label{eq.ev.op}
  \hat U(t)=\hat D^\dagger(\hat\beta)e^{-i\hat\omega t\hat a^\dagger\hat a}\hat D(\hat\beta)e^{i\eta t\hat\beta},
\end{align}
where $\hat\beta=\eta\hat\omega^{-1}$ and $\hat\omega=-\Delta_c+a_1N+a_2\hat J_x$, and $\hat D(\hat\beta)=e^{\hat\beta\hat a^\dagger-\hat\beta^\dagger\hat a}$
is a generalization of the displacement operator~\cite{knight_qo,scully1999quantum}. With Eq.~(\ref{eq.ev.op}), the operator $\hat h$ can be evaluated explicitly
(see Appendix \ref{app.der.gen} for details)
\begin{align}\label{eq.generator}
  \hat h&=\frac{\partial\hat\omega}{\partial\theta}\left(-i\frac{\hat\beta^2}{\eta}(\hat a^\dagger-\hat a)+t(\hat a^\dagger+\hat\beta)(\hat a+\hat\beta)+\right.\nonumber\\
  &\left.+i\frac{\hat\beta^2}{\eta}\left[(\hat a^\dagger+\hat\beta)e^{it\hat\omega}-(\hat a+\hat\beta)e^{-it\hat\omega}\right]+t\hat\beta^2\right).
\end{align}
A large QFI and thereby a high sensitivity, is achieved whenever $\hat h$ scales strongly, i.e., at least linearly, with the number of particles and the time $t$. This is
the case for the generator in Eq.~(\ref{eq.generator}), which contains terms scaling linearly with the number of qubits and photons, as well as with time $t$. To see it,
we rewrite $\hat h$ as
\begin{align}
  \hat h=t\frac{\partial\hat\omega}{\partial\theta}\hat a^\dagger\hat a+\hat f(\hat\omega,\eta,\hat a,\hat a^\dagger;t),
\end{align}
where the explicit form of $\hat f$ can be read-out from Eq.~(\ref{eq.generator}). 
In the absence of the drive, $\hat f$ is zero. In such a case, for a light-matter state
\begin{align}\label{eq.noon}
  \ket\psi=\frac{\ket{-\frac N2}+\ket{\frac N2}}{\sqrt2}\otimes\ket n,
\end{align}
which is composed of a superposition of eigenstates of $\hat J_x$ with the minimal and the maximal
eigenvalues ($N$-qubit cat state) and a photon Fock state, we obtain
\begin{align}\label{eq.heis.2}
  F_Q=t^2a_2'^2n^2N^2,
\end{align}
i.e., a Heisenberg scaling with both the number of qubits and photons~\cite{pezze2009entanglement}. Here and below, 
primes denote the derivatives of coefficients of the Hamiltonian (\ref{eq.ham}) over the parameter $\theta$.
The double-Heisenberg scaling of the QFI in \eqref{eq.heis.2} is a consequence of
cooperative effects: all the qubits are subject to the same effective magnetic
field whose strength is proportional to the number of photons.

Cooperative effects are present and enhance the sensitivity even
without resorting to non-classical states of light and entangled
states of the qubits.
Let us consider the tensor product of a coherent state of light $|\alpha\rangle$ and a coherent state of qubits, i.e.,  a state where all qubits point in the $z$ direction:
\begin{align}\label{eq.at.init}
  \ket{\psi_A}=\sum_{m=-\frac N2}^{\frac N2}C_m\ket m,\ \ C_m=\frac1{2^{\frac N2}}\sqrt{\binom N{\frac N2\pm m}},
\end{align}
where the sign $\pm$ depends on the choice of the direction along $z$
and $C_m$'s are the coefficients of the state in the basis of the
eigenstates $|m\rangle$ of $\hat J_x$.
For this state we have $\av{\hat J_x}=0$ and $\av{\hat J_x^2}=\frac{N}4$, thus 
\begin{align}\label{eq.qfi.class}
  F_Q=nt^2\Big[4\varphi'^2+(a_2')^2N(n+1)\Big],
\end{align}
where $n=|\alpha|^2$ and
\begin{align} \label{eq.static}
  \varphi=-\Delta_c+a_1N.
\end{align}
Though the QFI from Eq.~(\ref{eq.qfi.class}) is missing the double-Heisenberg scaling of Eq.~(\ref{eq.heis.2}), it still shows a Heisenberg scaling with the number of
qubits (since $\varphi$ scales with N) together with shot-noise scaling with the number of photons, or vice versa. The fact that this happens also without any quantum
correlations tells us that the Heisenberg scaling in this case is a classical cooperative effect where the dynamics in the estimation-parameter space is accelerated by a
factor proportional to the number of qubits or photons. An equivalent mechanism enhances the sensitivity of non-linear interferometers \cite{mitchell_2011}.

Finally, to go beyond the scaling $N^2n$ or $Nn^2$ with initially uncorrelated pure states of matter and light, and reach the double-Heisenberg scaling, when the QFI scales
as $N^2n^2$, the state requires to be at least entangled in qubit or nonclassical in photonic degrees of freedom. In the former case, the QFI contains
the term $\av{(\Delta \hat J_x)}n^2$, which yields the desired precision if the variance of the collective spin operator scales with $N^2$.
With the nonclassical photonic states, in the QFI the dominating term is $(a_1' N + a_2' \av{\hat J_x})^2\av{(\Delta\hat n)^2}$, which leads to very high precision
if the variance of the photonic distribution scales with $n^2$.

\subsection{Bounds for specific measurements}
\label{subsec:measurement_bounds}

Having found favorable scaling bounds for the sensitivity, one has
to determine which estimation strategies---that is, which measurement
observables and data processing protocols---allow to saturate those bounds.

In this section, we address this issue by considering the case where
the electromagnetic field is not driven. This simpler case is
generalized to the driven-dissipative case in the next section.
We specifically consider the bound given by Eq.~\eqref{eq.qfi.class},
which corresponds to the uncorrelated light-matter input state of  photonic coherent state $\ket\alpha$ (with the mean number of photons $n = \alpha^2$) 
and the coherent state of qubits given in Eq.~\eqref{eq.at.init}.

We first consider the case where the measurement is performed on
the qubits, specifically the $z$-component of the collective spin
operator. The simplest estimation strategy is to deduce $\theta$ 
from the mean value of the measurements of $\hat{J}_z$. It gives the well-known error propagation formula for the sensitivity
\begin{align}\label{eq.epf.at}
  \Delta^2\theta=\frac{\Delta^2\hat J_z}{\left(\frac{\partial\av{\hat J_z}}{\partial\theta}\right)^2}=\frac1{t^2}\frac1N\frac1{n(n+1)}\frac1{a_2'^2},
\end{align}
where the last equality is evaluated at optimal times such that $a_2t=k\times2\pi,\ k\in\mathbb N$.
(for the detailed derivation and a general formula valid for all times, see Appendix~\ref{app.atoms}.)
This sensitivity, due to the missing $\varphi'^2$ term, does not reach the the bound from Eq.~\eqref{eq.qfi.class}.
We thus conclude that, whenever the $\theta$-dependence of $a_2$ is stronger than the one of $\varphi$, most of the information about the parameter is 
accessible only with the qubit subsystem.
The estimation from the measurement of $\hat J_z$ is sensitive only to the dynamical qubit-induced phase shift of the mode frequency.

Let us now instead consider the case where the measurement is
performed on the photons via the quadrature operator \cite{knight_qo,scully1999quantum}
\begin{align}\label{eq.quad.def}
  \hat X_\phi=\frac12\left(\hat ae^{-i\frac\phi2}+\hat a^\dagger e^{i\frac\phi2}\right),
\end{align}
where $\phi$ is a phase that can be adjusted to maximize the signal. With help of Eq.~\eqref{eq.ev.op} and a coherent state of light at the input with
$\eta=0$, we obtain (see Appendix~\ref{app.photons} for details)
\begin{align}\label{eq.sens.quad}
  \Delta^2\theta=\frac{\Delta^2\hat X_\phi}{\left(\frac{\partial\av{\hat X_\phi}}{\partial\theta}\right)^2}=\frac1{t^2}\frac1{4n}\frac1{\varphi'^2},
\end{align}
again at optimal times $a_2t=2\pi k,\ k\in\mathbb N$ and with $\phi$ chosen such that $\sin^2(\varphi+\phi/2)=1$.  We see that a measurement performed on the photons
saturates the bound~\eqref{eq.qfi.class} if the contribution proportional to $a_2$ can be neglected. For these optimal times, the estimation of $\theta$ with the measurement of quadrature is
sensitive only to the static collective shift of the cavity frequency but insensitive to the dynamical shift.  

Therefore, given a classical input state of light and matter, by
performing the measurement on the qubits one can reach a sensitivity
scaling at the Heisenberg limit with the photon number and at the
shot-noise limit with the qubit number. If the measurement is performed
on the photons, the Heisenberg scaling is achieved with respect to the
number of qubits instead. 
This can be understood by the following reasoning. The estimation by measuring a subsystem is equivalent to averaging out over the remaining parts of the whole system.
Since the measured subsystem is described by a classical state, the precision cannot surpass the respective shot noise limit. The coefficient in the precision, however, is enhanced
due to the collective effects inherent to the Hamiltonian from Eq.~\eqref{eq.ham}.

\section{Impact of cavity pump and loss}
\label{sec:pump-loss}

In this section, we consider a more realistic case where the electromagnetic field is coherently driven, later addressing also the impact of the photon loss.

\subsection{Lossless case}

Starting from a vacuum state of the photons together with all qubits pointing in the $z$ direction, see Eq.~\eqref{eq.at.init}, the state at time $t$ is described by the
following density matrix
\begin{align}\label{eq.st}
  \hat\varrho(t)&=\sum_{m,m'}C_mC_{m'}\ketbra{\gamma_m}{\gamma_{m'}}\otimes\ketbra{m}{m'}\times\\
  &\times e^{i\eta(\beta_m-\beta_{m'})t}e^{-i[\beta_m^2\sin(\omega_m t)-\beta_{m'}^2\sin(\omega_{m'} t)]}\nn,
\end{align}
where $\ket{\gamma_m}$ denotes a coherent state of light with the amplitude 
$\gamma_m=\beta_m(e^{-i\omega_mt}-1)$, with  $\omega_m=-\Delta_c+a_1N+a_2m$ and $\beta_m={\eta}/{\omega_m}$ (see Appendix~\ref{app.evolution}).

The state of the light, tracing out the subspace of qubits, is a mixture of coherent states
\begin{align}\label{eq.st.light}
  \hat\varrho_L(t)=\tr{\hat\varrho(t)}_A=\sum_mC^2_m\ketbra{\gamma_m}{\gamma_m}.
\end{align}
We note that the average number of photons is given by
\begin{align}\label{eq.mean.ph}
  n=\av{\hat n}=\tr{\hat n\hat\varrho_L(t)}=\sum_mC^2_m|\gamma_m|^2.
\end{align}
Depending on the relative strength of the parameters entering the Hamiltonian and the properties of the state of the system, we can specify two different limits:
coherent and incoherent regime. Below, we address these in more details.

\subsubsection{Coherent regime}\label{subsec.model.pure}

For small times, the impact of the dynamical phase shift on the dynamics is negligible. In such a case, $\omega_m$ is independent of the state of the matter and is given
only by the static shift of the cavity frequency, i.e., $\omega_m \approx \varphi$., see Eq.~\eqref{eq.static}. The requirement is that the following
condition
\begin{align}\label{eq.approx}
  |-\Delta_c+a_1N|\gg|a_2|m
\end{align}
is satisfied for all $m$ that significantly contribute to the state in Eq.~\eqref{eq.st}.

The state remains in this coherent regime, as long as the time $t$ is sufficiently 
short so that the amplitude $C_m$ with maximal $m$'s that significantly contribute to the state, 
i.e., with $m = \pm\sqrt N$, 
has approximately the same phase.
This is true up to $t\simeq\tau_c=\frac{2}{\sqrt N \vert a_2\vert}$. Within this
time-frame we have $\gamma_m\simeq\gamma=\frac\eta\varphi(e^{-i\varphi t}-1)$
for all $m$ and the sum Eq.~\eqref{eq.st.light} can be explicitly calculated,
giving a pure coherent state of light $\hat\varrho_L(t)\simeq\ketbra{\gamma}{\gamma}$.
Consequently, the number of photons oscillates as
\begin{align}\label{eq.rev}
  \av{\hat n}\simeq|\gamma|^2=2 \bar n \sin^2\left(\frac{\varphi t}2\right)
\end{align}
where $\bar n=2\frac{\eta^2}{\varphi^2}$ is the number of photons
averaged over one oscillation period.
When the time $t$ exceeds $\tau_c$, contributions to Eq.~(\ref{eq.mean.ph}) oscillate out-of-phase, giving
$\av{\hat n}\simeq\bar n.$
Time-oscillations of the mean photon number emerge again when $t\simeq\tau_r=\frac\pi{a_2}$, giving a pattern of  collapses and revivals, in analogy to the
dynamics of a two-level atom within the Jaynes--Cummings model, driven by a monochromatic coherent state of light~\cite{knight_qo,scully1999quantum}.

We now focus on this oscillatory regime and calculate the sensitivity
using the estimation strategies discussed in Section~\ref{subsec:measurement_bounds}.
Let us first consider the measurement of the light quadrature, for
which the sensitivity, calculated again with the error propagation formula reads
\begin{align}\label{eq.epf.ph2}
  \Delta^2\theta=\frac{\av{(\Delta\hat X_{\phi})^2}}{\left(\frac{\partial\av{\hat X_{\phi}}}{\partial\theta}\right)^2}  \simeq\frac1{t^2}\frac1{2\bar n}\frac1{\varphi'^2}
\end{align}
with the phase chosen such that $\phi+2\varphi t=(2k+1)\pi,\ k\in\mathbb N$ (see Appendix~\ref{app.sens} for details). 
We used Eq.~\eqref{eq.st.light} to get
\begin{align}\label{eq.quad.mean}
  \av{\hat X_\phi}=\re{\gamma e^{-i\frac\phi2}},\ \ \ &\av{(\Delta\hat X_{\phi})^2}\equiv\av{\hat X^2_\phi}-\av{\hat X_\phi}^2=\frac14,
\end{align}
where $\re{\cdot}$ stands for the real part.
We see that in the driven case, the measurement of the quadrature in
the coherent oscillatory regime gives the same sensitivity as
predicted by using an input coherent photon state with amplitude set
by $\eta/\varphi$. Here, since $\omega_m\approx \varphi$, the dynamical frequency shift does not significantly modify the state, 
and the information about the parameter is encoded in the static shift of the cavity frequency.

We now turn to the measurement of the qubits $\hat{J}_z$. We use the
Heisenberg equations of motion for the collective spin operators
\begin{align}\label{eq.heis.app}
  \partial_t\hat J_{z/y}(t)=-i[\hat J_{z/y}(t),\hat H]=\pm a_2\hat J_{y/z}(t)\hat n(t).
\end{align}
In the oscillatory regime, when light is in a pure coherent
state, we approximately replace $\hat n(t)$ with the average number of
photons, i.e. $\partial_t\hat J_{z/y}(t)\simeq \pm a_2\hat J_{y/z}(t)|\gamma|^2$.
This gives $\hat J_z(t)=\hat J_z\cos(\chi)+\hat J_y\sin(\chi)$, with 
\begin{equation}\label{eq.chi}
  \chi \equiv a_2\int_0^t\!\!d\tau\,|\gamma|^2=\bar n(1-\sinc(\varphi t))a_2t.
\end{equation}
The error propagation formula then yields 
\begin{align}\label{eq.epf.at}
  \Delta^2\theta=\frac{\Delta^2\hat J_z(t)}{\left(\partial_\theta\av{\hat J_z(t)}\right)^2}=\frac1{(\chi')^2}\frac1N\simeq\frac1{t^2}\frac1N\frac1{\bar n^2}\frac1{a_2'^2},
\end{align}
if $\vert\frac{a_2}{a_2'}\frac{\varphi'}{\varphi}\vert\ll1$ and  $\sinc(\varphi t)\ll1$.
Note that, although the oscillations of the photonic dynamics revive
periodically, the mean-field approximation used above can be safely applied only once. This is 
because in the long collapse periods, though the dynamics of the photonic population is virtually frozen, the atomic operators undergo a complex dynamics, setting an unknown initial condition
for the solution in the next oscillatory regime.
Also for this estimation strategy, within the coherent oscillatory
regime the sensitivity coincides with the one predicted with the
proper input coherent state with sufficiently large number of photons.

For times larger than $\tau_c$, the photonic dynamics is frozen so we
do not expect the $t^{-2}$ scaling of the sensitivity encountered in
the oscillatory case (see Eq.~\eqref{eq.epf.ph2}). 
Indeed, the mean quadrature and its 
variance are now
\begin{align}
  \av{\hat X_\phi}=-\sqrt{\frac{\bar n}2}\cos\left(\frac\phi2\right),\ \ \ \av{(\Delta\hat X_\phi)^2}=\frac14(\bar n+1).
\end{align}
thus
\begin{align}\label{eq.sens.col}
  \Delta^2\theta\simeq\frac1{2\bar n}\frac{\varphi^2}{\varphi'^2},
\end{align}
when $\bar n\gg1$. Here, the inverse scaling with time as well as with the number of qubits is lost, due to presence of $\varphi^2$ in the
numerator.  For the case where the measurement is performed on the qubits, an analytical calculation similar to that presented in
Eqs~\eqref{eq.heis.app}-\eqref{eq.epf.at} is not possible after $\tau_c$, as light is not in a pure coherent state anymore. Therefore, we must rely on the numerical exact
diagonalization of the Hamiltonian~\eqref{eq.ham} which gives a sensitivity which is orders of magnitude smaller than in the oscillatory regime.

\subsubsection{Incoherent regime}\label{subsec.model.mix}

When the impact of the dynamical phase shift due to the presence of qubits cannot be neglected, 
the state of the photons cannot be described by a single coherent state. 
In such a case, when the condition in Eq.~\eqref{eq.approx} is not satisfied, the replacement of the mixture in Eq.~\eqref{eq.st.light} with a pure 
coherent state is not justified at all times, and
the mean number of photons is given with the general formula from Eq.~\eqref{eq.mean.ph}. The  first two moments of the quadrature are now
\begin{subequations}\label{eq.quad.mix}
  \begin{align}
    &\av{\hat X_\phi}=\sum_mC_m^2\re{\gamma_me^{-i\frac\phi2}},\\
    &\av{\hat X^2_\phi}=\sum_mC_m^2\re{\gamma_me^{-i\frac\phi2}}^2+\frac14.
  \end{align}
\end{subequations}
Although one has to resort to numerical simulations in this general case, we show that in presence of the photon loss,
the sensitivity~\eqref{eq.epf.ph2} can be still determined even
in the incoherent regime.

\begin{figure}[t!]
  \includegraphics[width=\columnwidth]{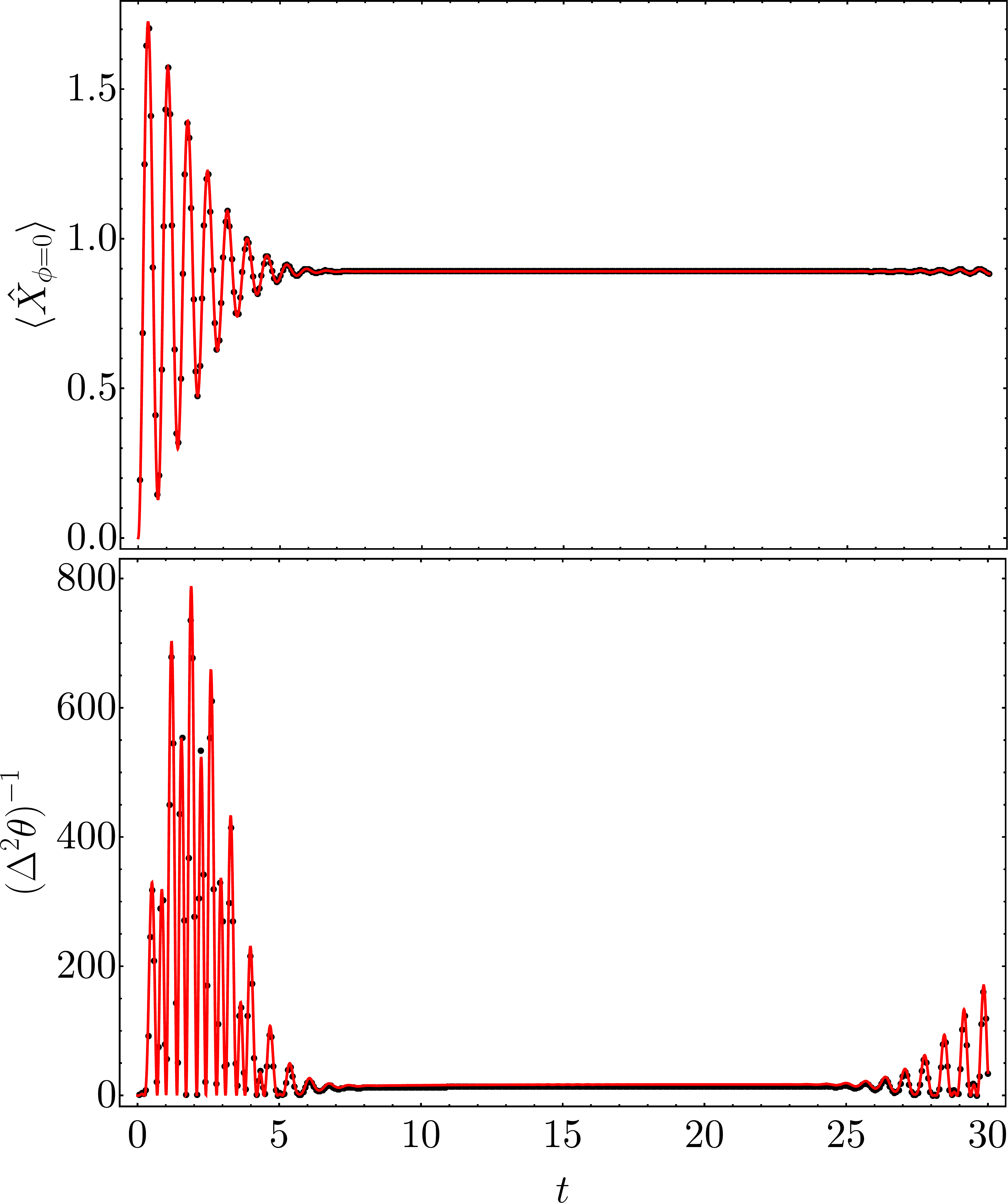}
  \caption{Photons. The average value of the quadrature $\hat X_\phi$ for $\phi=0$ (top) and inverse of the error propagation formula (bottom) for quadrature as
    a function of time $t$. Black points represent results of numerical calculations, while red solid line stands for the analytic solution.  The parameters are: $N=20$,
    $a_1 = -0.5$, $a_2= -0.2$, $\Delta_c = -1$, $a_1' = a_2'= 1$, $\eta=8$ and $\kappa = 0.3$, while the initial state consists of a vacuum state of the photons together with all qubits pointing in the $z$ direction.}\label{fig.light}
\end{figure}

\begin{figure}[t!]
  \includegraphics[width=\columnwidth]{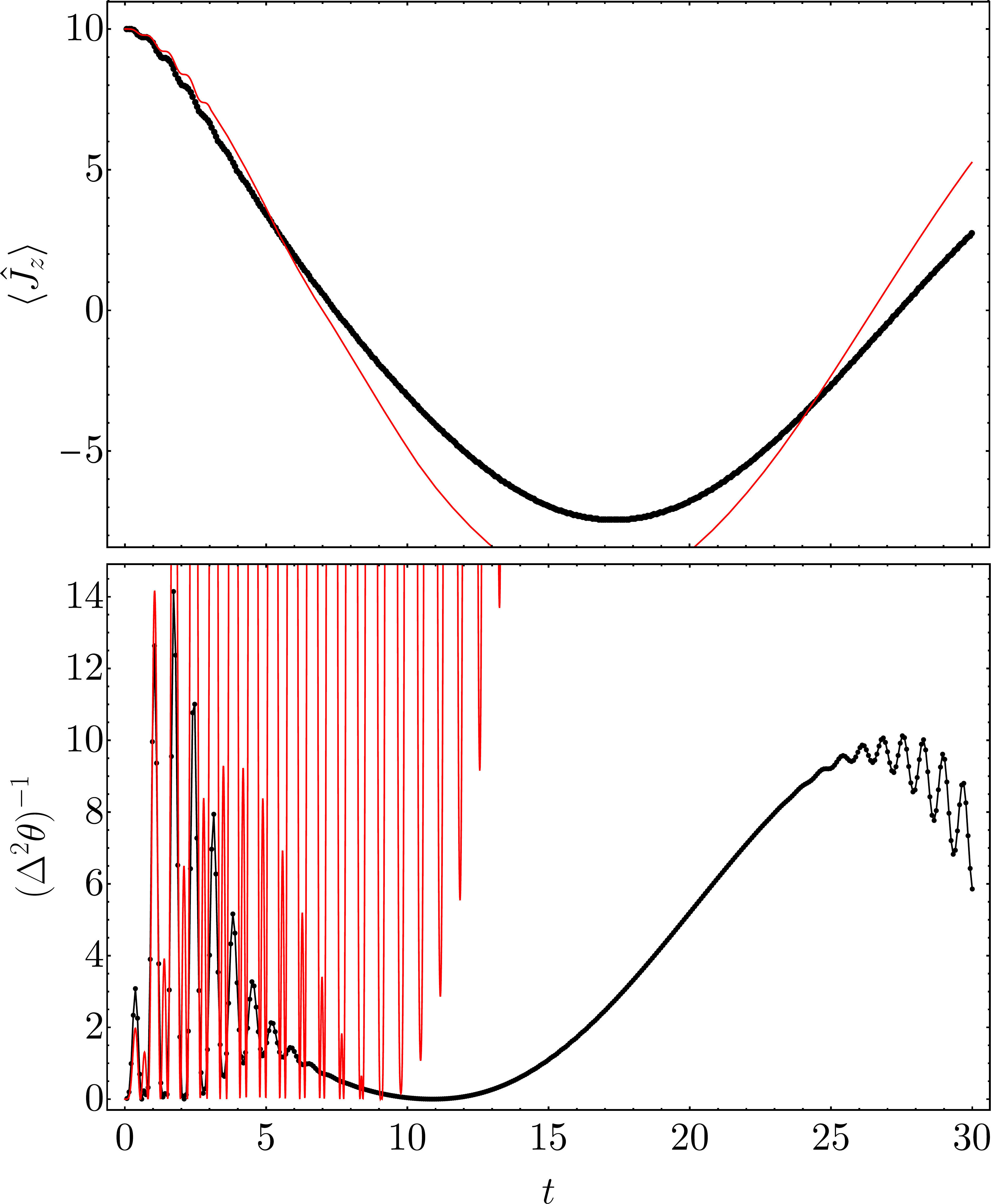}
  \caption{Qubits. The average value of the operator $\hat{J}_z$ (top) and inverse of the error propagation formula (bottom) for $\hat J_z$  as a function
    of time $t$. Black points represent results of numerical calculations, while red solid line stands for the analytic solution valid when $t<\tau_c = 2.24$.
    Parameters used for calculations are the same as in Fig.~\ref{fig.light}.}\label{fig.atoms}
\end{figure}

\subsection{Impact of photon losses}\label{sec.losses}
In this section we include the possibility for photons to be lost from
the electromagnetic mode at a rate $\kappa$. The dynamics of the
system is then described by the following quantum master equation \cite{breuer2002theory} for the density
matrix of the system:
\begin{align}
  \frac{d}{dt}\hat{\varrho}=-\frac{i}{\hbar}[\hat{H},\hat{\varrho}]+\kappa\left(\hat{a}\hat{\varrho}\hat{a}^\dag-\frac12
  \{\hat{a}^\dag\hat{a},\hat{\varrho}\}\right).
\end{align}

To proceed, we again distinguish separate the coherent and the incoherent dynamics regime, according to the condition from Eq.~\eqref{eq.approx}.

\subsubsection{Coherent regime}\label{subsec.losses.pure}

In the coherent regime, when $t\lesssim\tau_c$, 
we model the photon dynamics by effectively including the loss term in the equation for the coherent amplitude, i.e.,
$\partial_t\gamma=\left(-i\varphi-\frac{\kappa}{2}\right)\gamma-i\eta.$
With the solution of the photonic state, which is given by
\begin{align}\label{eq.gamma.loss}
  \gamma=\frac{\eta}{\varphi-i\frac{\kappa}{2}}\left(e^{-i\varphi t-\frac{\kappa}{2} t}-1\right),
\end{align}
we determine the mean and the variance of the quadrature by inserting $\gamma$ from Eq.~\eqref{eq.gamma.loss} into 
Eq.~\eqref{eq.quad.mean}.
In the short-time limit $\kappa t\ll 1$, we obtain the following sensitivity:
\begin{align}\label{eq.sens.loss}
  \Delta^2\theta\simeq\frac1{2\bar n_\kappa(\varphi't)^2}\frac{\varphi^2+\frac{\kappa^2}4}{\varphi^2},
\end{align}
where $\bar n_\kappa=2\frac{\eta^2}{\varphi^2+\frac{\kappa^2}{4}}$ is
the time-averaged number of photons. In the opposite limit, when $\kappa t\gg1$, but still $t\lesssim\tau_c$, we have
$\gamma\simeq-\frac{\eta}{\varphi-i\frac{\kappa}{2}}.$
This gives the sensitivity from the mean quadrature:
\begin{align}
  \Delta^2\theta=\frac1{\bar n_\kappa\varphi'^2}\frac{(\varphi^2+\frac{\kappa^2}4)^3}{(\varphi^2-\frac{\kappa^2}4)^2}.
\end{align}
Similarly to Eq.~\eqref{eq.sens.col}, the presence of $\varphi^2$ in the numerator neutralizes the scaling of $\varphi'^2$ with the number
of qubits and thus the collective effect is absent.

Adapting the approach from Eqs.~\eqref{eq.heis.app}--\eqref{eq.epf.at}
to the presence of photon loss, we determine the sensitivity from the measurement of $\hat J_z(t)$, see Eq.~\eqref{eq.chi}:
\begin{align}\label{eq.chi.loss}
  &\chi=\frac12a_2\bar n_\kappa\Bigg[t+\frac{1-e^{-\kappa t}}\kappa-\frac{1}{\varphi^2+\frac{\kappa^2}{4}}\times\\
    &\times\left(\kappa-e^{-\frac{\kappa t}2}\kappa\cos(\varphi t)+2e^{-\frac{\kappa t}2}\varphi\sin(\varphi t)\right)\Bigg]\nn.
\end{align}
When $\kappa t\ll1$, the error propagation formula reproduces Eq.~\eqref{eq.epf.at} with $\bar n$ replaced by $\bar n_\kappa$, namely
\begin{align}\label{eq.sens.loss2}
  \Delta^2\theta\simeq\frac1{t^2}\frac1N\frac1{\bar n_\kappa^2}\frac1{a_2'^2}.
\end{align}

In the limit $\kappa t\rightarrow\infty$, we obtain the sensitivity 
\begin{align}\label{eq.sens.loss2}
  \Delta^2\theta\simeq\frac1{t^2}\frac1N\frac1{\bar n_\kappa^2}\frac1{(\frac{a'_2}2-\frac{\varphi\varphi'}{\varphi^2+\frac{\kappa^2}4})^2}.
\end{align}
The solutions presented here are compared with numerical calculations on Figs.~\ref{fig.light} and~\ref{fig.atoms}. To illustrate the usefulness of the formulas we
derived, we take a vacuum state of the photons together with $N=20$ qubits pointing in the $z$ direction,   $a_1 = -0.5$, $a_2= -0.2$, $\Delta_c = -1$, $a_1' = a_2'= 1$, $\eta=8$ and $\kappa = 0.3$, so that both oscillations and collapse are visible. In this case the
important time scale is given by $\tau_c = 2.24$. Estimation from the mean quadrature agrees perfectly with analytical expression presented in this section,
recovering both collapse and revival. On the other hand, the estimation from the qubits deviates once the initial oscillations are repressed, which is when~$t>\tau_c$.

The results of Eqs.~\eqref{eq.sens.loss} and~\eqref{eq.sens.loss2}, 
show that the collective scalings of the sensitivity with the number of photons and qubits can be retained in the presence of losses for both estimation strategies.

\subsubsection{Incoherent regime}\label{subsec.losses.mix}

We now turn to the incoherent regime where the
condition~\eqref{eq.approx} does not hold. In this case, we solve the same
equation for the coherent amplitude $\gamma_m$ as above, this time in each subspace of fixed $m$.
We obtain
\begin{align}
  \gamma_m=\frac{\eta}{\omega_m-i\frac\kappa2}\left(e^{-i\omega_mt-\kappa t}-1\right).
\end{align}
Although an analytical expression for the sensitivity is not available
in general, a closed formula for $\Delta\theta$ from the quadrature
measurement can be found in some regimes, which provides insight into the scalings.

First, taking the limit of large times $\kappa t\gg1$ and assuming that
$\kappa$ can be neglected in comparison to $\omega_m$ for all $m$, we get
$
  \gamma_m=-{\eta}/{\omega_m}.
$
We can now calculate the mean number of photons using
Eq.~\eqref{eq.mean.ph}, and similarly the two lowest moments of the
quadrature with Eqs.~\eqref{eq.quad.mix}, yielding $\av{\hat
  X_\phi}=-\cos\left(\phi/2\right)\sum_mC_m^2\eta/{\omega_m}$
and $\av{\hat X^2_\phi}=1/4+\cos^2\left(\phi/2\right)\av{\hat n}$.
Now, if $\av{\hat n}\gg1$ and $\omega_m\ll \eta$ for those
$m$'s where $C_m$ are significantly non-zero,  we have $\av{\hat X_\phi}\ll\av{\hat n}$ and $\av{\hat X^2_\phi}\simeq\cos^2\left(\phi/2\right)\av{\hat n}$,
and thus $\Delta^2\hat X\simeq\cos^2\left(\phi/2\right)\av{\hat n}$. 
The error propagation formula then yields the shot-noise scaling with the photon number and the enhanced scaling with the number of qubits:
\begin{align}\label{eq.sens.limit}
  \Delta^2\theta=
  \frac{\av{\hat n}}{\left(\sum_mC_m^2\frac\eta{\omega_m^2}\omega_m'\right)^2} \approx \frac{\eta^2}{N^2a_1'^2\av{\hat n}},
\end{align}
where in the last step we approximated: $\omega_m'=a_1'N-a_2'm \approx a_1'N$.

In the next section, we use our results
to calculate the sensitivity of the estimation of the gravitational acceleration in a realistic setting.

\section{Application to gravimetry}\label{sec.app}

Here we offer a concrete example where the cooperative
enhancement of the sensitivity can be exploited to measure precisely a
fundamental constant in a realistic experimental setup.

\begin{figure}[t]
  \includegraphics[width=\columnwidth]{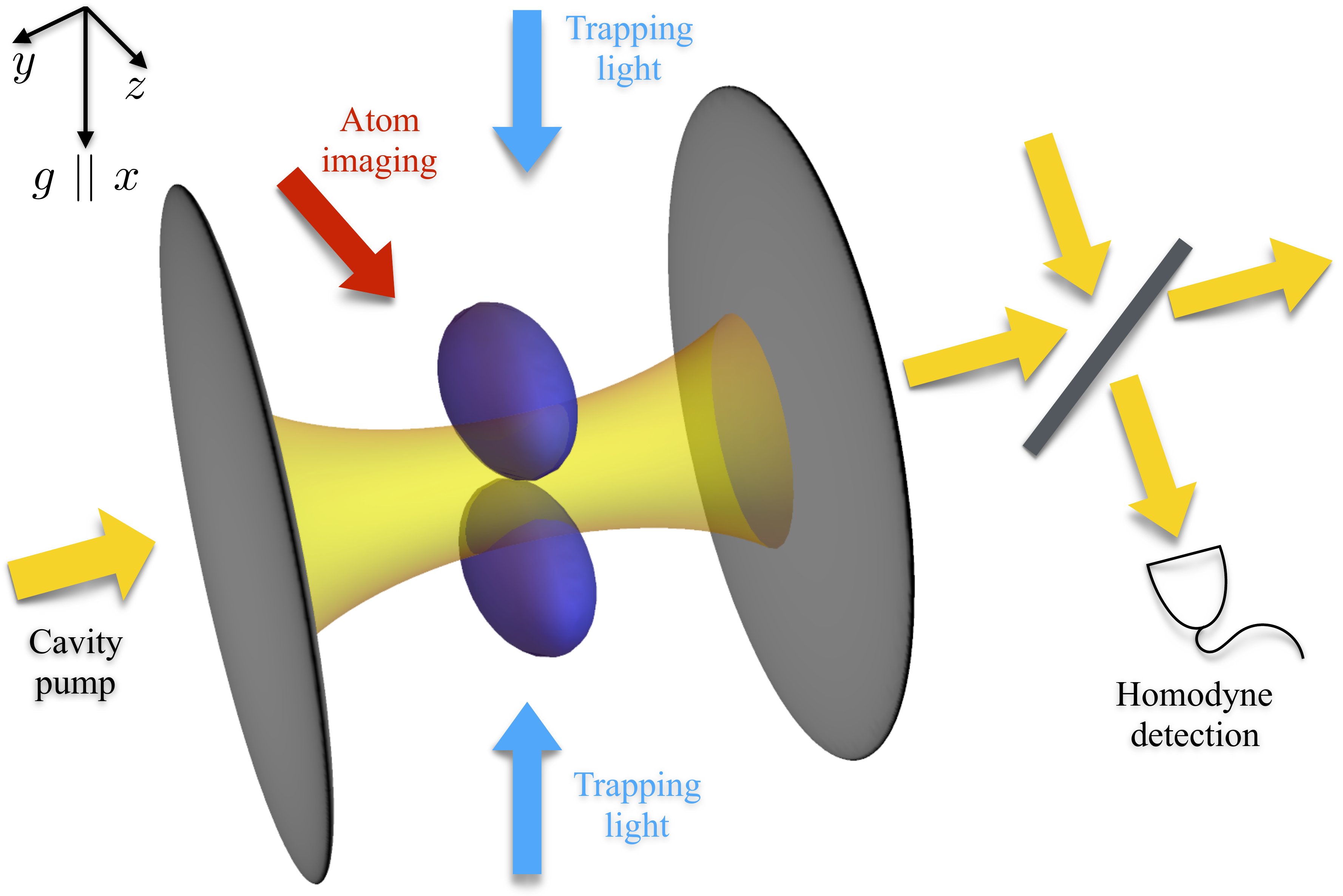}
  \caption{The scheme of a hybrid light-matter system used as a gravitational sensor. The standing wave (yellow beam) of the cavity formed between two mirrors (gray)
    modifies the tunneling barrier between the two wells (blue) formed by the trapping light (blue arrows).  The cavity is driven by an external laser (see the main text
    in Sec.~\ref{sec.app}), and the outgoing light can be analyzed in a homodyne detector. Alternatively, the atoms can be monitored by an auxiliary laser (red arrow).}
  \label{fig.setup}
\end{figure}

Specifically, we consider an optical cavity with resonance
frequency $\omega_c$, driven by a laser with a strength $\eta$ and frequency $\omega_l$, 
far detuned from an electronic transition of atoms (with resonance
frequency $\omega_a$), i.e., $\Delta_a=\omega_l-\omega_a$ is by far the largest scale, so that the excited state can be adiabatically
eliminated. The atoms are assumed to form a BEC trapped in a
double-well potential. We consider the configuration show in
Fig.~\ref{fig.setup} where the standing wave of the cavity modifies
the tunneling barrier between the two wells (the classical dynamics of
such system has been studied in \cite{szirmai}). The Hamiltonian of
the system can be mapped to our model Hamiltonian \eqref{eq.ham}
(see Appendix~\ref{app.ham}), where the characteristic photon frequency becomes the
cavity detuning from the laser: $\omega_0\to\Delta_c = \omega_l -
\omega_c$. The qubit collective spin operators are expressed in terms
of the bosonic operators $\hat b_{1,2}$ annihilating an atom in the
potential well $1,2$:
$\hat J_x=\frac12(\hat b_1^\dagger\hat b_2+\hat b_1\hat b^\dagger_2)$,
$\hat J_y=\frac1{2i}(\hat b_1^\dagger\hat b_2-\hat b_1\hat
b^\dagger_2)$, and $\hat J_z=\frac12(\hat b_1^\dagger\hat b_1-\hat
b^\dagger_2\hat b_2)$, together with $\hat N=\hat b_1^\dagger\hat b_1+\hat b^\dagger_2\hat b_2$.
The coefficients $a_1$ and $a_2$ of our Hamiltonian \eqref{eq.ham},
which in this example quantify the ac-Stark shift and the cavity
assisted tunneling constant, respectively, are expressed in terms of
the overlap integrals \cite{szirmai}
\begin{align}\label{eq.ints}
  I_{ij}=\frac{\hbar U_0(1 + e^{-k^2l^2_H})}{2L\pi\sigma\sqrt{l_H^2 + \sigma^2}}\int dx w^*_i(x)w_j(x)e^{-x^2/\sigma^2},
\end{align}
through $a_1=I_{11}+I_{22}$ and $a_2=I_{12}+I_{21}$.
Here $L$ is the cavity length, $w_1(x)$ and $w_2(x)$ are the
Wannier-like atomic wave-functions centered around the two minima of the
double-well potential, $l_H$ is the characteristic length
of the strong harmonic confinement, $k$ is the wavevector of the
cavity light, $\sigma$ is the cavity beam waist, $U_0 =
\frac{\Omega_R^2}{\Delta_a}$ is the dispersive shift of the cavity
frequency per atom, and $\Omega_R$ 
is cavity-mode Rabi frequency quantifying the light-matter coupling. 

We want to propose this hybrid light-matter system as a precise
gravitational sensor exploiting the cooperative effects. The linear
gravitational potential $V_{\rm grav}(x)=gx$, see Fig.~\ref{fig.setup}, acts by
shifting the double-well potential with respect to the cavity
axis. This has a two-fold impact on the system. First, it modifies the
Hamiltonian parameters through the integrals from Eq.~\eqref{eq.ints} by shifting
the gaussian beam profile $e^{-x^2/\sigma^2}\rightarrow e^{-(x-x_0)^2/\sigma^2}$, where $x_0=g/\omega_x^2$. Second, it adds an energy-imbalance term $\delta\hat J_z$ to the Hamiltonian, where
\begin{align}
  \delta=\int dx\,V_{\rm grav}(x)\left(|w_1(x)|^2-|w_2(x)|^2\right).
\end{align}

In order to provide a realistic estimate for the sensitivity of the measurement of $g$, we use the following parameters.  We take a cavity with finesse $\mathcal{F} =
4\times 10^6$, length $L = 2.954$ mm, and loss rate $\kappa = 25$ kHz.  Setting the distance between the wells $D = 1.6$ $\mu$m, the characteristic length of the harmonic
oscillator $a_{ho} = 0.8$ $\mu$m and characteristic length in the perpendicular direction $l_H = 1.6$ $\mu$m gives the trap frequencies 
$(\omega_x, \omega_y, \omega_z) = 2\pi \times (181, 41, 41)$ Hz.  
We choose $^{87}$Rb atoms and the detuning from the atomic transition $\Delta_a=- 2.015$ GHz.  The width of the TEM$_{00}$ mode
function is $\sigma=13.65$ $\mu$m, which finally gives $a_1=-171$~kHz and $a_2=-103$~kHz. The derivatives of these two coefficients with respect to the
metrological parameter $g$ are $a_1'=13.6$ $\frac{\text{GHz}}{\text{m}}$ and $a_2'=8.34$ $\frac{\text{GHz}}{\text{m}}$.

With $N \approx 9.22 \times 10^5$ atoms, the renormalized cavity detuning $\varphi$ can be tuned to $-400$ Hz (the bare detuning being $\Delta_c=-155$ GHz), giving the
mean number of photons $\bar{n} \simeq 5.6\times 10^4$ for $\eta=100$ MHz. If $t> 1/\kappa$, and when the input state of atoms is the
coherent spin state [see Eq.~\eqref{eq.at.init}], the formula from Eq.~\eqref{eq.sens.limit} yields the precision $\Delta g = 8 \times 10^{-7}g$. Such sensitivity can be
reached within a measurement time on the order of $\kappa$:
\begin{align}
  \frac{\Delta g}{g}=5\times10^{-9} \sqrt{\mathrm{Hz}}.
\end{align}

\section{Conclusions}\label{sec.conc}

We have shown that a hybrid system of matter and light can act as a sensing device in which the cooperative effects play a prominent role.
These effects generically enhance the precision by improving the
scaling with the number of particles in both subsystems.

By considering a fundamental model of $N$ qubits coupled to a single
electromagnetic mode, we showed that the precision in estimating the
light-matter coupling constant exhibits a double-Heisenberg scaling $\Delta\theta \propto 1/(Nn)$, where $n$ is the number of photons. This scaling requires the use of an entangled state of
matter or a nonclassical state of photons. However, even for classical states a Heisenberg scaling with the number of qubits or photons can be reached.

To illustrate the usefulness of our hybrid light-matter sensor, we proposed a specific, experimentally feasible scheme in which a Bose-Einstein condensate is trapped in a
double-well potential within an optical cavity.  We predicted that, even
taking into account photon loss, the sensor can determine the gravitational acceleration $g$
with a relative precision reaching $\Delta g/g\simeq10^{-9}\text{Hz}^{-1/2}$. Such a precision, which still can be improved by employing nonclassical states of matter
and light, is comparable to the one predicted for a supersolid state
of atoms in a optical cavity~\cite{gietka_ss_grav_2019}.

\section{Acknowledgements}
AN and JC are supported by Project no. 2017/25/Z/ST2/03039, funded by the National Science Centre, Poland, under the QuantERA programme.

\appendix

\section{Hamiltonian and the coefficents}\label{app.ham}

We outline the derivation of the Hamiltonian of the coupled atom-light system, which is discussed in full extent in Ref.~\cite{szirmai}. We consider
an ultra-cold gas of $N$ two-level bosons, trapped in a double-well potential immersed in an optical cavity of length $L$. 
The cavity is pumped  with a monochromatic radiation of frequency $\omega_l$, which is far detuned from the frequency of the internal atomic transition, 
allowing for an adiabatic elimination of the excited state. Atoms occupy only the low-lying pair of degenerate states of the 
double-well potential $V_{\rm dw}(x)$, so the atomic
field is described by two operators $\hat b_{\mathrm 1/2}$, which annihilate a boson in the right/left site of the trap, i.e.,
\begin{align}
  \hat\Psi(x)=w_1(x)\hat b_1+w_2(x)\hat b_2.
\end{align}
Here, $w_{1/2}(x)$ are the Wannier-like states localized in the corresponding site for the trap.

Combining the two-mode model for atoms and a single-mode description of the photonic field, we obtain the Hamiltonian, which is a sum of the free Hamiltonian of light (l), atoms (a) and an
interaction part (a+l)
\begin{align}\label{app.eq.ham.p2}
  \hat H=\hat H_{\rm l}+\hat H_{\rm a}+\hat H_{\rm a+l},
\end{align}
where
\begin{subequations}
  \begin{align}
    &\hat H_{\rm l}=-\Delta_c\hat n+\eta(\hat a+\hat a^\dagger)\\
    &\hat H_{\rm a}=\omega_J\hat J_x+\delta\hat J_z\\
    &\hat H_{\rm a+l}=(a_1\hat N+\tilde a_1\hat J_z+a_2\hat J_x)\hat n.
  \end{align}
\end{subequations}
Here $\hat a$ is the photonic annihilation operator and $\hat n=\hat a^\dagger\hat a$. The angular momentum operators are
\begin{subequations}
  \begin{align}
    &\hat J_x=\frac12(\hat b_1^\dagger\hat b_2+\hat b_2\hat b^\dagger_1),\\
    &\hat J_y=\frac1{2i}(\hat b_1^\dagger\hat b_2-\hat b_2\hat b^\dagger_1),\\
    &\hat J_z=\frac1{2}(\hat b_1^\dagger\hat b_1-\hat b^\dagger_2\hat b_2).
  \end{align}
\end{subequations}
and $\hat N=\hat b_1^\dagger\hat b_1+\hat b_2^\dagger\hat b_2$ is the atom-number operator. 
The $\Delta_C = \omega_l - \omega_c$ is the cavity detuning ($\omega_c=\frac{2\pi c}L$) and $\eta$ is the strength of the pump.

The coefficients $a_1$, $\tilde a_1$ and $a_2$ (the symmetric and the anti-symmetric part of the ac-Stark shift and the cavity assisted tunneling constant, respectively), are expressed in terms of 
\begin{align}
  I_{ij}=\frac{\hbar U_0(1 + e^{-k^2l^2_H})}{2L\pi\sigma\sqrt{l_H^2 + \sigma^2}}\int dx w^*_i(x)w_j(x)e^{-x^2/\sigma^2},
\end{align}
as $a_1=I_{11}+I_{22}$, $\tilde a_1=I_{11}-I_{22}$ and $a_2=I_{12}+I_{21}$.
Here, $l_H$ is the length
of the strong harmonic confinement, $k$ is the wavevector of the cavity light, $\sigma$ is beam waist at close to the middle of the cavity, $U_0 = \frac{\Omega_R^2}{\Delta_a}$, and $\Omega_R$ 
is single mode Rabi frequency and $\Delta_a = \omega_l - \omega_a$ the detuning of the laser from the atomic transition of frequency $\omega_a$. 

Finally, the parameter determining the atom-only Hamiltonian are the bare Josephson energy and the energy imbalance between the wells induced by some external potential $V_{\rm ext}(x)$.
\begin{subequations}
  \begin{align}
    &\omega_J=-2\int dxw_1(x)\left(-\frac{\hbar^2}{2m}\frac{d^2}{dx^2}+V_{\rm dw}(x)\right)w_2(x),\\
    &\delta=\int dxV_{\rm ext}(x)\left(w^2_1(x)-w^2_2(x)\right)
  \end{align}
\end{subequations}
As argued in the main text, for realistic parameters $\delta$ and $\tilde a_1$ can be neglected leaving the Hamiltonian in the form
\begin{align}\label{app.eq.ham}
  \hat H=-\Delta_c\hat n+\eta(\hat a+\hat a^\dagger)+(a_1\hat N+a_2\hat J_x)\hat n.
\end{align}
where the free atomic term $\omega_J\hat J_x$ was included as a phase factor in the dynamics of the initial state. In a fixed-$N$ subspace the operator $\hat N$ is replaced with $N$.

\section{Evolution operator}\label{app.evolution}

We now derive the expression for the evolution operator. The Hamiltonian from Eq.~\eqref{app.eq.ham} can be written as
\begin{align}
  \hat H=\hat\omega\hat a^\dagger\hat a+\eta(\hat a+\hat a^\dagger),
\end{align}
where $\hat\omega=a_1N-\Delta_c+a_2\hat J_x$. Now we observe that
\begin{align}
  \hat H=\hat\omega\hat D^\dagger(\hat\beta)\hat a^\dagger\hat a\hat D(\hat\beta)-\eta\hat\beta,
\end{align}
where $\hat\beta=\eta\hat\omega^{-1}$, while 
\begin{align}
  \hat D(\hat\beta)=e^{\hat\beta\hat a^\dagger-\hat\beta^\dagger\hat a}
\end{align}
is the generalized displacement operator. Since $[\hat\omega,\hat\beta]=0$, we can write the evolution operator as follows
\begin{align}\label{app.eq.ev.op}
  \hat U(t)=\hat D^\dagger(\hat\beta)e^{-i\hat\omega t\hat a^\dagger\hat a}\hat D(\hat\beta)e^{i\eta t\hat\beta},
\end{align}
The initial state has a general form
\begin{align}\label{app.eq.st.0}
  \hat\varrho(0)=\sum_{n,n'=0}^\infty\sum_{m,m'=-\frac N2}^{\frac N2}\varrho_{nn'}^{mm'}\ketbra{n,m}{n',m'},
\end{align}
where $\ket{n,m}$ denoted a photonic Fock state and an eigenstate of the atomic operator $\hat J_x$, namely
\begin{align}
  \ket{n,m}=\ket n\otimes\ket m,\ \ \ \hat a^\dagger\hat a\ket n=n\ket n,\ \hat J_x\ket m=m\ket m.
\end{align}
The action of the evolution operator (\ref{app.eq.ev.op}) on the density matrix from Eq.~(\ref{app.eq.st.0}) gives
\begin{align}\label{app.eq.dens.evo}
  \hat\varrho(t)&=\sum\limits_{\substack{n,n'\\m,m'}}\varrho_{nn'}^{mm'}\hat D^\dagger(\beta_m)e^{-i\omega_mt\hat a^\dagger\hat a}\hat D(\beta_m)e^{i\eta(\beta_m-\beta_{m'})t}\times\nn\\
  &\times\ketbra{n,m}{n',m'}
  \hat D^\dagger(\beta_{m'})e^{i\omega_{m'}t\hat a^\dagger\hat a}\hat D(\beta_{m'}),
\end{align}
where $\gamma_m=\beta_m(e^{-i\omega_mt}-1)$, $\omega_m=-\Delta_c+a_1N+a_2m$ and $\beta_m=\frac{\eta}{\omega_m}$.
Note that
\begin{align}\label{app.eq.disp}
  &\ket{\Phi_1}\equiv\hat D(\beta)\ket n=\frac1{\sqrt{n!}}\hat D(\beta)(\hat a^\dagger)^n\ket0=\\
  &\frac1{\sqrt{n!}}\hat D(\beta)(\hat a^\dagger)^n\hat D^\dagger(\beta)\hat D(\beta)\ket0=\frac1{\sqrt{n!}}(\hat a^\dagger-\beta)^n\ket\beta,\nn
\end{align}
as $\beta\in\mathbb R$. With this expression at hand, we can take the next step and act with the free-evolution term
\begin{align}
  \ket{\Phi_2}&\equiv e^{-i\omega\hat a^\dagger\hat a t}\ket{\Phi_1}=\frac1{\sqrt{n!}}e^{-i\omega\hat a^\dagger\hat a t}(\hat a^\dagger-\beta)^n\ket\beta\nn\\
  &=\frac1{\sqrt{n!}}e^{-i\omega\hat a^\dagger\hat a t}(\hat a^\dagger-\beta)^ne^{i\omega\hat a^\dagger\hat a t}e^{-i\omega\hat a^\dagger\hat a t}\ket\beta\nn\\
  &=\frac1{\sqrt{n!}}(\hat a^\dagger e^{-i\omega t}-\beta)^n\ket{\beta e^{-i\omega t}}.
\end{align}
In the last step, we add the second displacement operator, to get
\begin{align}
  &\hat D^\dagger(\beta)\ket{\Phi_2}=\frac1{\sqrt{n!}}\hat D^\dagger(\beta)(\hat a^\dagger e^{-i\omega t}-\beta)^n\ket{\beta e^{-i\omega t}}\nn\\
  &=\frac1{\sqrt{n!}}\hat D^\dagger(\beta)(\hat a^\dagger e^{-i\omega t}-\beta)^n\hat D(\beta)\hat D^\dagger(\beta)\ket{\beta e^{-i\omega t}}\nonumber\\
  &=\frac1{\sqrt{n!}}e^{-i\beta^2\sin(\omega t)}((\hat a^\dagger+\beta)e^{-i\omega t}-\beta)^n\ket{\beta e^{-i\omega t}-\beta}\nn\\
  &=\frac1{\sqrt{n!}}e^{-i\beta^2\sin(\omega t)}(\hat a^\dagger e^{-i\omega t}+\gamma)^n\ket{\gamma},
\end{align}
where $\gamma=\beta(e^{-i\omega t}-1)$. We again use the displacement operator
\begin{align}
  &(\hat a^\dagger e^{-i\omega t}+\gamma)^n\ket{\gamma}=\frac1{\sqrt{n!}}(\hat a^\dagger e^{-i\omega t}+\gamma)^n\hat D(\gamma)\ket{0}\nn\\
  &=\hat D(\gamma)\hat D^\dagger(\gamma)(\hat a^\dagger e^{-i\omega t}+\gamma)^n\hat D(\gamma)\ket{0}\nonumber\\
  &=\hat D(\gamma)((\hat a^\dagger+\gamma^*) e^{-i\omega t}+\gamma)^n\ket{0}.
\end{align}
But note that
\begin{align}
  \gamma^*e^{-i\omega t}+\gamma&=\beta(e^{i\omega t}-1)e^{-i\omega t}+\gamma\nn\\
  &=\beta(1-e^{-i\omega t})+\gamma=-\gamma+\gamma=0.
\end{align}
Therefore, we obtain the final expression
\begin{align}
  &\hat D^\dagger(\beta)e^{-i\omega\hat a^\dagger\hat a t}\hat D(\beta)\ket n=\frac{e^{i\beta^2\sin(\omega t)}}{\sqrt{n!}}\hat D(\gamma)(\hat a^\dagger e^{-i\omega t})^n\ket{0}\nn\\
  &=e^{-in\omega t}e^{-i\beta^2\sin(\omega t)}\hat D(\gamma)\ket{n}.
\end{align}
We now plug this result into Eq.~(\ref{app.eq.dens.evo}) and obtain 
\begin{align}\label{app.eq.st}
  \hat\varrho(t)&=\sum_{m,m'}C_mC_{m'}\ketbra{\gamma_m}{\gamma_{m'}}\otimes\ketbra{m}{m'}\nn\\
  &\times e^{i\eta(\beta_m-\beta_{m'})t}e^{-i[\beta_m^2\sin(\omega_m t)-\beta_{m'}^2\sin(\omega_{m'} t)]}
\end{align}
as used in the main text.

\section{Derivation of the generator $\hat h$ from Eq~(\ref{eq.generator})}\label{app.der.gen}
The generator of the interferometric/metrological transformation is equal to
\begin{align}
  \hat h=i(\partial_\theta\hat U)\hat U^\dagger.
\end{align}
The derivative over the parameter will hit all the parameter-dependent parts of the evolution operator. For instance
\begin{align}
  \partial_\theta\hat\beta=-\eta\hat\omega^{-2}\frac{\partial\hat\omega}{\partial\theta}=-\frac{\hat\beta^2}\eta\frac{\partial\hat\omega}{\partial\theta}.
\end{align}
All other steps leading to Eq.~(\ref{eq.generator}) follow immediately from the properties of the displacement operator.

\section{Sensitivities in the $\eta=0$ case}
We now separately consider the no-pump case where initially light is in a coherent state $\ket\alpha$, and derive the expressions for the error propagation formula for atoms- and photons-only.
The complete density matrix in such case is given by
\begin{align}\label{eq.full.nopump}
  \hat\varrho(t)=\sum\limits_{m,m'}\varrho^{(A)}_{m,m'}\ketbra{\gamma_m}{\gamma_{m'}}\otimes\ketbra{m}{m'},
\end{align}
where $\gamma_m=\alpha e^{-i\omega_mt}$ (note that $\varrho^{(A)}_{m,m'}=C_mC_{m'}$, so $\hat\varrho(t)$ is pure). However the density-matix representation is useful for the claculation of the reduced
matrices. This is the starting point for the discussion in the remaining part of this Appendix.

\subsection{Error propagation formula for atoms}\label{app.atoms}
We first calculate the atomic density matrix by tracing-out the photonic degree of freedom. We obtain
\begin{align}\label{eq.at}
  \hat\varrho_A&=\tr{\hat\varrho(t)}_L=\sum_{m,m'=0}^N\varrho^{(A)}_{m,m'}\braket{\gamma_{m'}}{\gamma_m}\ketbra{m}{m'}=\nn\\
  &=\sum_{m,m'=0}^N\varrho^{(A)}_{m,m'}e^{-\alpha^2[1-\cos(\delta(m-m'))]}\times\nn\\
  &\times e^{i\alpha^2\sin(\delta(m-m'))}\ketbra{m}{m'},
\end{align}
where $\delta=a_2t$. To calculate the error propagation formula, we note that
\begin{align}
  \hat J_z\ket{m}=\frac12&\left(\sqrt{\left(\frac N2+m+1\right)\left(\frac N2-m\right)}\ket{m+1}\right.\nn\\
  &\left.+\sqrt{\left(\frac N2+m\right)\left(\frac N2-m+1\right)}\ket{m-1}\right)
\end{align}
and analogically for $\hat J_z^2$. Therefore we obtain
\begin{subequations}
  \begin{align}
    &\av{\hat J_z}=\frac N2e^{n(\cos\delta-1)}\cos(n\sin\delta)\\
    &\av{\hat J^2_z}=\frac N8(N-1)\Big[e^{n(\cos2\delta-1)}\cos(n\sin2\delta)+1\Big]+\frac N4\\
    &\partial_\theta\av{\hat J_z}=
    -\frac N2na_2'te^{n(\cos\delta-1)}\sin\left[\delta+n\sin\delta\right].
  \end{align}
\end{subequations}
These expression, plugged into the error propagation formula (\ref{eq.epf.at}) gives
\begin{align}
  \Delta^2\theta=\frac1{Nn^2(a_2't)^2}\frac{\Delta^2\hat J_z}{\frac14e^{2n(\cos\delta-1)}\sin^2\left[\delta+n\sin\delta\right]}.
\end{align}
This can be optimized by setting $\delta=k\times2\pi,\ k\in\mathbb N$, which gives Eq.~(\ref{eq.epf.at}).

\subsection{Error propagation formula for photons}\label{app.photons}
We now take the state from Eq.~(\ref{eq.full.nopump}) and trace-out the atomic degree of freedom to obtain
\begin{align}\label{eq.at}
  \hat\varrho_L&=\tr{\hat\varrho(t)}_A=\sum_{m=0}^N\varrho^{(A)}_{m,m}\ketbra{\gamma_{m}}{\gamma_m},
\end{align}
i.e., the state is an incoherent mixture of coherent states. From this representation of the photonic state, we immediately obtain
\begin{align}
  \av{\hat X}&=\frac12\sum_mC_m^2\left(\gamma_me^{-i\frac\phi2}+\gamma^*_me^{i\frac\phi2}\right)=\nn\\
  &=\frac\alpha2\sum_mC_m^2\left(e^{-i(\omega_mt+\frac\phi2)}+\gamma^*_me^{i(\omega_m t+\frac\phi2)}\right)=\nn\\
  &=\alpha\cos\left(\varphi t+\frac\phi2\right)\cos^N\left(\frac\delta2\right),
\end{align}
where in the last step we used the explicit expression for $C_m$ from Eq.~(\ref{eq.at.init}).
In a similar fashion, we obtain
\begin{align}
  \av{\hat X^2}&=\frac14+\sum_mC_m^2(\gamma_m^2e^{-i\phi}+\gamma_m^{*2}e^{i\phi}+2|\gamma_m|^2)=\nn\\
  &=\frac14+\frac{\alpha^2}2+\frac{\alpha^2}2\cos(2\varphi t+\phi)\cos^N(\delta).
\end{align}
for the mean of its square.

From these two results, the variance of $\hat X$ can be obtained and optimized (i.e., minimized) with respect to $\delta$. By picking $\delta=k\times2\pi,\ k\in\mathbb N$, we obtain
\begin{align}
  \av{(\Delta\hat X)^2}=\frac14
\end{align}
and the error propagation formula gives the sensitivity equal to
\begin{align}\label{eq.sens.quad}
  \Delta^2\theta=\frac1{t^2}\frac1{4n}\frac1{\varphi'^2\sin^2(\varphi+\phi/2)}.
\end{align}
Once we set $\sin^2(\varphi+\phi/2)=1$, we recover Eq.~\eqref{eq.sens.quad}.

\section{Sensitivities for $\eta\neq0$}\label{app.sens}

The photonic quadrature is
\begin{align}\label{app.eq.quad.def}
  \hat X_\phi=\frac12\left(\hat ae^{-i\frac\phi2}+\hat a^\dagger e^{i\frac\phi2}\right),
\end{align}
and using Eq.~\eqref{eq.quad.mix} and Eq.~\eqref{app.eq.st} we obtain the mean and the mean square
\begin{subequations}\label{app.eq.quad.mean}
  \begin{align}
    &\av{\hat X_\phi}=\sum_mC_m^2\frac\eta{\omega_m}\left[\cos\left(\omega_mt+\frac\phi2\right)-\cos\left(\frac\phi2\right)\right]\\
    &\av{\hat X_\phi^2}=\sum_mC_m^2\frac{\eta^2}{\omega^2_m}\left[\cos\left(\omega_mt+\frac\phi2\right)-\cos\left(\frac\phi2\right)\right]^2+\frac14.
  \end{align}
\end{subequations}
In the oscillatory regime and when the approximation~\eqref{eq.approx} holds, the depndence of  $\omega_m$ on $m$ can be dropped, giving
\begin{subequations}
  \begin{align}
    &\av{\hat X_\phi}\simeq\left[\cos\left(\varphi t+\frac\phi2\right)-\cos\left(\frac\phi2\right)\right]\frac\eta{\varphi}.\\
    &\av{\hat X_\phi^2}\simeq\av{\hat X_\phi}^2+\frac14.
  \end{align}
\end{subequations}
The sensitivity is inversely proportional to the square of the derivative of $\av{\hat X_\phi}$, equal to
\begin{align}
  \frac{\partial{\av{\hat X_\phi}}}{\partial\theta}=-\frac\eta{\varphi^2}\varphi'\Big[&\cos\left(\varphi t+\frac\phi2\right)-\cos\left(\frac\phi2\right)\nn\\
    &+\sin\left(\varphi t+\frac\phi2\right)\varphi t\Bigg].
\end{align}
thus by choosing $\phi$ in such a way that $\varphi t+\frac\phi2=\frac\pi2+k\pi,\ k\in\mathbb N$, 
we obtain
\begin{align}\label{app.eq.epf.ph2}
  \Delta^2\theta\simeq\frac1{t^2}\frac1{2\bar n}\frac1{\varphi'^2}.
\end{align}

For atoms, the mean-field approximation described in the main text gives with 
\begin{align}
  \Delta^2\theta=\frac{\Delta^2\hat J_z(t)}{\left(\partial_\theta\av{\hat J_z(t)}\right)^2}=\frac1N\frac1{(\chi')^2},
\end{align}
where 
\begin{align}
  \chi(t)=a_2\int_0^t\!\!d\tau\,|\gamma|^2=2a_2t\frac{\eta^2}{\varphi^2}(1-\sinc(\varphi t)).
\end{align}
For those  sufficiently late instants of time $t$, when $\sinc(\varphi t)\ll1$, the error propagation formula gives
\begin{align}
  \Delta^2\theta=\frac{\Delta^2\hat J_z(t)}{\left(\partial_\theta\av{\hat J_z(t)}\right)^2}=\frac1{t^2}\frac1N\frac1{\bar n^2}\frac1{(a_2')^2}\frac1{(1-2\frac{a_2}{a_2'}\frac{\varphi'}{\varphi})^2}.
\end{align}
In the collapse regime, the cosine functions cancel out in Eqs~\eqref{app.eq.quad.mean}, while the cosine squared averages to $1/2$, giving
\begin{subequations}
  \begin{align}
    \av{\hat X_\phi}&\simeq-\frac\eta{\varphi}\cos\left(\frac\phi2\right)\\
    \av{\hat X_\phi^2}&\simeq\frac{\eta^2}{\varphi^2}\left[\frac12+\cos^2\left(\frac\phi2\right)\right]+\frac14.
  \end{align}
\end{subequations}
The variance is bigger than in the oscillatory regime and the mean does grow with time. The sensitivity at $\phi=0$ is
\begin{align}
  \Delta^2\theta=\frac{\frac12\frac{\eta^2}{\varphi^2}+\frac14}{\frac{\eta^4}{\varphi^4}\frac{\varphi'^2}{\varphi^2}}\simeq\frac1{\bar n\frac{\varphi'^2}{\varphi^2}},
\end{align}
which is $2t^2\varphi^2$ worse than Eq.~\eqref{app.eq.epf.ph2}.

\end{document}